\documentclass[aps,pra,floatfix,groupedaddress,superscriptaddress,notitlepage,showpacs,nofootinbib]{revtex4-1}
\usepackage{graphicx,graphics,epstopdf,color,times,bm,bbm,dsfont,amssymb,amsmath,amsfonts,hyperref,natbib}

\newcommand{\Ket}[1]{|#1\rangle\hskip-0.5mm\rangle}

\newcommand{\Bra}[1]{\langle\hskip-0.5mm\langle#1|}

\newcommand{\BraKet}[2]{\langle\hskip-0.5mm\langle#1|#2\rangle\hskip-0.5mm\rangle}

\newcommand{\ignore}[1]{}

\let\oldsqrt\sqrt
\def\sqrt{\mathpalette\DHLhksqrt}
\def\DHLhksqrt#1#2{%
\setbox0=\hbox{$#1\oldsqrt{#2\,}$}\dimen0=\ht0
\advance\dimen0-0.2\ht0
\setbox2=\hbox{\vrule height\ht0 depth -\dimen0}%
{\box0\lower0.4pt\box2}}

\newcommand{\ba}[1]{\begin{array}{#1}}
\newcommand{\ea}{\end{array}}

\newcommand{\be}{\begin{equation}}
\newcommand{\ee}{\end{equation}}
\newcommand{\bea}{\begin{eqnarray}}
\newcommand{\eea}{\end{eqnarray}}
\newcommand{\beann}{\begin{eqnarray*}}
\newcommand{\eeann}{\end{eqnarray*}}

\newtheorem{definition}{Definition}[section]
\newtheorem{proposition}{Proposition}[section]

\newtheorem{remark}{Remark}[section]

\begin{document}
\title{Dissipative quantum metrology in manybody systems of identical particles}
\author{F. Benatti}
\affiliation{Dipartimento di Fisica, Universit\`{a} di Trieste, I-34151 Trieste, Italy}
\affiliation{Istituto Nazionale di Fisica Nucleare, Sezione di Trieste, I-34151 Trieste, Italy}
\author{S. Alipour}
\affiliation{Department of Physics, Sharif University of Technology, Tehran, Iran}
\author{A. T. Rezakhani}
\affiliation{Department of Physics, Sharif University of Technology, Tehran, Iran}

%%%%%%%%%%%%%%%%%%%%%%%%%%%%%%%%%%%%%%%%%%%%%%%%%%%%%%%%%%%%%%%
\begin{abstract}
Estimation of physical parameters is a must in almost any part of science and technology. The enhancement of the performances in this task, e.g., beating the standard classical shot-noise limit, using available physical resources is a major goal in metrology. Quantum metrology in closed systems has indicated that entanglement in such systems may be a useful resource. However, it is not yet fully understood whether in open quantum systems such enhancements may still show up. Here, we consider a dissipative (open) quantum system of identical particles in which a parameter of the open dynamics itself is to be estimated. We employ a recently-developed dissipative quantum metrology framework, and investigate whether the entanglement produced in the course of the dissipative dynamics may help the estimation task. Specifically, we show that even in a Markovian dynamics, in which states become less distinguishable in time, at small enough times entanglement generated by the dynamics may offer some advantage over the classical shot-noise limit.
\end{abstract}
\maketitle
%%%%%%%%%%%%%%%%%%%%%%%%%%%%%%%%%%%%%%%%%%%%%%%%%%%%%%%%%%%%%%%
\section{Introduction}

Quantum entanglement is perhaps the most intriguing consequence of the linear structure of quantum mechanics. Since the EPR gedankenexperiment \cite{einapr1935}, the perception of the phenomenon changed from being an argument against the compatibility of quantum mechanics with special relativity to becoming a physical resource usable to perform tasks that would be impossible in a completely classical world. Entangled states have now applications in as many different fields as quantum communications, quantum computation \cite{Nielsen-Chuang:book} and, lately, in quantum metrology \cite{meter}.

In the latter case, entangled $N$-qubit states have been proposed as means to beat the so-called shot-noise limit accuracy
in parameter estimation~\cite{Kitagawa1993,Wineland1994}. Indeed, if a parameter $\alpha$ appears in a unitary transformation $\mathrm{e}^{-i\alpha J}$ generated by a self-adjoint operator $J$, one can measure $\alpha$ by subjecting a system in an initial state $\varrho$ to the unitary operator $\mathrm{e}^{-i\alpha J}$, mapping it into $\mathrm{e}^{-i\alpha J}\varrho\mathrm{e}^{i\alpha J}$. The quantum Cram\'er-Rao bound (QCRB), in the form $\delta\alpha\geqslant 1/\sqrt{\mathcal{F}[\varrho,J]}$, provides a lower limit to the accuracy of estimation $\delta \alpha$ in terms of the inverse of the square of the so-called quantum Fisher information (QFI) \cite{Braunstein,Luo,Hel,Hol,den,paris}, $\mathcal{F}[\varrho,J]$, associated with the generator of the unitary transformation and the state of the system. Now, if $\varrho$ is a separable state, the QFI scales as $O(N)$ with the number of particles in the system, $N$, while it may scale faster for entangled $\varrho$s.

Ultracold atoms in optical lattices are among the most successful physical setups for studying quantum manybody effects in condensed-matter physics \cite{Bloch05} and quantum information \cite{Lewenstein2,Amico08}. More recently, these systems have been proposed to be employed in experimental realizations of quantum metrological protocols \cite{Giov04,Oberthaler2}. First steps in this direction have recently been realized: entangled states in systems of ultra-cold atoms have been generated through spin-squeezing techniques \cite{Oberthaler1,Treutlein}, with the aim of using them as input states in interferometric apparatuses, specifically constructed for quantum-enhanced metrological applications. In such systems, the initial $N$-qubit states are rotated by means of collective spin operators. For instance, the self-adjoint operator $J$ above may be chosen to be $J_z=(1/2)\sum_{j=1}^N\sigma^{(j)}_z$, that is the $z$ component of the total angular momentum of the system, with $\sigma^{(j)}_z$ the $z$ Pauli matrix of the $j$th qubit. In the case of distinguishable qubits, all these rotations are local in the sense that they split into the tensor product of the single-qubit unitary operators: $\mathrm{e}^{-i\alpha J}=\bigotimes_{j=1}^{N} \mathrm{e}^{-i\alpha\sigma^{(j)}_z/2}$. Therefore, preliminary spin squeezing of separable states is necessary in order to introduce (nonlocal) quantum correlations, whereby opening the possibility of sub-shot-noise errors in parameter estimation.

However, while quantum entanglement has a clear formulation in relation to distinguishable particle systems \cite{horrhh2009}, the entanglement of identical particles is still lacking an agreed-upon status \cite{Zanardi1,Cirac,ghigmw2002,Bruss,GhiMar,Zanardi3,Narnhofer,Viola1,Viola2,benffm2010,Abu,Marmo,benffm2012,bala}.
In the case of trapped ultra-cold atoms, the qubits involved are identical---a fact that has often not been fully appreciated in recent literature on quantum metrology. Indeed, the indistinguishability of the particles asks for a rethinking of definitions and concepts originally introduced in the case of distinguishable qubits.

More specifically, the usual notion of separability is strictly associated with the tensor product structure of the Hilbert space, $\mathbb{H}_N=\bigotimes_{j=1}^N\mathbb{H}_j$, which is natural in the case of $N$ distinguishable particles. On the other hand, pure bosonic states must be symmetric under exchange of particles, and mixed states must be convex combinations
of projections onto such states. This fact necessitates a different approach to the notions of nonlocality and entanglement based not on a structure related to the particle aspect of first quantization, but rather on the behavior of correlation functions of commuting observables more generally related to the mode description typical of second quantization picture.

In this spirit, a generalized notion of entanglement, based on the mode description, has been introduced in Ref.~\cite{benffm2010}, which reduces to the standard definition for distinguishable qubits. In particular, in the mode description of a system consisting of $N$ identical particles, the lack of a definite tensor product structure in the corresponding Bose or Fermi algebra forces the notion of locality of observables to be expressed in terms of their commutativity with respect to other observables. One can then extend the notion of separable states by referring to whether the mean value of the product of two observables, each belonging to one of a pair of commuting subalgebras $(\mathcal{A},\mathcal{B})$, can be expressed as a convex combinations of products of the mean values of the single observables. This generalization has been proved to be relevant for metrological applications involving identical bosons \cite{benffmJPB,BeBr}.

In order to be of use in practical metrological applications, the entanglement present in the state of the system subjected to the experimental protocol must be protected against decoherence effects coming from the system being weakly (but not negligibly)
coupled to an external reservoir. In other words, especially when delicate properties as entanglement are concerned, quantum systems ought to be considered as open with a time evolution generated---under some conditions---by a master equation that incorporates noise and dissipation induced by the ambient environment. Note that dissipation and noise inherent in an open quantum dynamics are not always sources of only decoherence; sometimes, the structure of the master equation might as well allow for the creation of quantum correlations \cite{BFP}. This is most easily seen in a second-quantized context \cite{ABFM}. Indeed, separability and entanglement are defined relative to a pair of commuting subalgebras $(\mathcal{A},\mathcal{B})$, therefore the effects of any time evolution, be it reversible (noiseless and dissipation-free) or irreversible (dissipative and noisy), may destroy nonlocality with respect to the pair $(\mathcal{A},\mathcal{B})$, but create nonlocal correlations with respect to another commuting pair $(\mathcal{C},\mathcal{D})$ obtained from the former by, e.g., a Bogolubov transformation.

It is known that if the open quantum dynamics is described by a semigroup of completely positive, trace-preserving (CPT) dynamical maps $\Gamma_t$, the QFI cannot but decrease \cite{den}. Strictly speaking, given the one-parameter family of states $\varrho_t=\Gamma_t[\varrho]$ generated by the CPT dynamical maps $\Gamma_t$ out of the initial state $\varrho$, the QFI relative to the measurement of the parameter $\alpha$ in
$\mathrm{e}^{-i\alpha J}$ by using the rotated states $\mathrm{e}^{-i\alpha J}\varrho_t\mathrm{e}^{i\alpha J}$ cannot increase in time, namely $0\leqslant s\leqslant t$ implies $\mathcal{F}[\varrho_t,J]\leqslant \mathcal{F}[\varrho_s,J]$.

Rather, in the following, we study a different scenario, in which the parameter to be estimated is inscribed directly in the generator of the open quantum dynamics itself and not superimposed by an external unitary operator as $\mathrm{e}^{-i\alpha J}$. In particular, we shall consider the dissipative time evolution of an open quantum system consisting of $N$ two-mode bosons governed by an entanglement-generating master equation, and investigate whether the generated entanglement is useful to go beyond the shot-noise limit. To this end, we need first to develop a nonunitary framework for the QFI and related techniques.

%%%%%%%%%%%%%%%%%%%%%%%%%%%%%%%%%%%%%%%%%%%%%%%%%%%%%%%%%%%%%%%
\section{$N$ two-mode boson dissipative time evolution}

As a test model for the issues raised in the introduction, we shall consider a simple phase-damping continuous process affecting two bosonic modes $\mathrm{a}$ and $\mathrm{b}$. Their dissipative time evolution is governed by the master equation
\begin{equation}
\partial_t\varrho(t)=\gamma\big(J_z\varrho(t)\,J_z-\frac{1}{2}
\big\{J_z^2,\varrho(t)\big\}\big)\equiv L_\gamma[\varrho],
\label{1}
\end{equation}
where $\varrho$ is a density matrix for this two-mode bosonic system, e.g., a system of $N$ bosons with modes corresponding to being in the left and right well of a double-well potential, and
\begin{equation}
J_z=(\mathrm{a}^\dag\mathrm{a}-\mathrm{b}^\dag\mathrm{b})/2,\quad [\mathrm{a},\mathrm{a}^\dag]=[\mathrm{b},\mathrm{b}^\dag]=\openone,\quad [\mathrm{a},\mathrm{b}^\dag]=0.
\label{2}
\end{equation}
The solution to Eq.~\eqref{1} can easily be found by considering the orthonormal basis in the $(N+1)$--dimensional
$N$-boson sector given by
\be
\label{6}
\vert k,N-k\rangle\equiv \frac{(\mathrm{a}^\dag)^{k}\,(\mathrm{b}^\dag)^{(N-k)}}{\sqrt{k!(N-k)!}}\vert 0\rangle\ ,\quad k=0,1,\ldots,N,
\ee
which are the eigenstates of $J_z$ with the eigenvalues $(2k-N)/2$. Setting
\begin{equation}
\varrho_{k\ell}=\langle k,N-k\vert\varrho\vert \ell,N-\ell\rangle,
\label{entries}
\end{equation}
it follows that
\be
\partial_{t}\varrho_{k\ell}(t)=-\frac{\gamma}{2}(k-\ell)^2\varrho_{k\ell}(t)\Longrightarrow \varrho_{k\ell}(t)=
{\rm e}^{-\gamma t(k-\ell)^2/2}\varrho_{k\ell}.
\label{ME}
\ee
A representation-independent expression for the solution can be obtained by writing the exponential as
\begin{equation}
\label{expon}
{\rm e}^{-\gamma t(k-\ell)^2/2}=\frac{1}{2\sqrt{\pi}}\int_{-\infty}^{+\infty}{\rm d}u\,{\rm e}^{-u^2/4}\,
{\rm e}^{-i\sqrt{\gamma t/2}u(k-\ell)}\ ,
\end{equation}
whence
\begin{equation}
\varrho(t)=\frac{1}{2\sqrt{\pi}}\int_{-\infty}^{+\infty}{\rm d}u~\mathrm{e}^{-u^2/4}\mathrm{e}^{-i\sqrt{\gamma t/2} u J_z}\varrho~\mathrm{e}^{i\sqrt{\gamma t/2}u J_z}.
\label{3}
\end{equation}
The maps $\Gamma_t:\varrho\mapsto\varrho(t)$ form a semigroup and are completely-positive unital maps for they are expressed in a continuous Kraus-Strinespring form.

%%%%%%%%%%%%%%%%%%%%%%%%%%%%%%%%%%%%%%%%%%%%%%%%%%%%%%%%%%%%%%%
\section{Two-mode boson entanglement}

As outlined in the introduction, when dealing with identical particles, a convenient setting is provided by second quantization whereby, instead of asking about the existence of nonlocal correlations between particles, one should
inspect the entanglement of different modes associated with creation and annihilation operators of orthogonal single-particle states \cite{benffm2010}. We shall denote by $\mathcal{M}$ the algebra generated by the operators \footnote{As bosonic creation and annihilation operators are unbounded, the standard technique \cite{BraRob,BraRob2} is to consider their exponentiation to unitary Weyl operators and the $C^*$-algebra generated thereby.}.

In a second-quantized setting, the usual bipartite entanglement of distinguishable particles turns into the entanglement with respect to a local bipartition $(\mathcal{A},\mathcal{B})$. In the case of two bosonic modes $\mathrm{a}$ and $\mathrm{b}$, $\mathcal{A}$ and $\mathcal{B}$ are the commuting subalgebras generated by these operators. Commutativity expresses algebraic independence. Hence, the local character of an operator, which for two particles
is embodied by tensor products $A\otimes B$, can be algebraically extended to all operators of the form $AB$ with $A\in\mathcal{A}$ and $B\in\mathcal{B}$, with the two subalgebras forming a commuting pair $(\mathcal{A},\mathcal{B})$. Note that this is true also in the standard setting where $A\otimes\openone$ and $\openone\otimes B$ belong to the commuting subalgebras.

\begin{definition}
A local bipartition of the boson algebra $\mathcal{M}$ is any pair of commutinig subalgebras $(\mathcal{A},\mathcal{B})$
contained in $\mathcal{M}$ such that $\mathcal{A}$ and $\mathcal{B}$ generate $\mathcal{M}$.
An operator $M\in\mathcal{M}$ is local with respect to $(\mathcal{A},\mathcal{B})$ if $M=AB$ where $A\in\mathcal{A}$ and $B\in\mathcal{B}$.
\label{def0}
\end{definition}

In the following, by state of the two-mode $N$-boson system under consideration we shall mean any density matrix acting on the $(N+1)$-dimensional Hilbert space spanned by the orthogonal occupation number eigenstates \eqref{6}. Thus, the standard formulation of separable states naturally generalizes to a bipartition-dependent one as follows:

\begin{definition}
A state $\varrho$ on the $N$ boson sector is  $(\mathcal{A},\mathcal{B})$-separable if and only if for all $A\in\mathcal{A}$ and $B\in\mathcal{B}$
\be
\label{sep}
\mathrm{Tr}[\varrho\,AB]=\sum_j\lambda_j {\rm Tr}[\varrho^{(1)}_j\,A]\, {\rm Tr}[\varrho^{(2)}_j\,B]\ ,
\ee
where $\lambda_j\geqslant 0$ are weights such that $\sum_j\lambda_j=1$, and $\varrho^{(1,2)}_j$ are other density matrices on the $N$-boson sector of the Fock space.
\label{def1}
\end{definition}

\begin{remark}
This definition of separable states is nothing but the second-quantized formulation of the one valid for distinguishable particles.
Indeed, when a bipartite density matrix $\varrho_{AB}$ can be written as a convex combination $\varrho_{AB}=\sum_j\lambda_j\varrho_j^{A}\otimes\varrho^{B}_j$ of density matrices of the first and second parties, the mean value of all local observables of the form  $A\otimes B$ split into the convex combination of the mean values of the observables $A$ and $B$ with respect to the local density matrices. In a second quantized formalism there is no tensor product structure available but only local subalgebras that commute and states that evaluate the mean values of the corresponding observables.
\label{seprem}
\end{remark}

It turns out that in the case of $N$ two-mode boson systems, the partial transposition with respect to one of the modes in the representation given by the orthonormal basis \eqref{6} is an exhaustive entanglement witness as for two qubits or one qubit and one qutrit \cite{benffm2010}.

\begin{proposition}
\label{prop1}
A mixed state $\varrho$ of an $N$ two-mode boson system is $(\mathcal{A},\mathcal{B})$-separable in the sense of Definition \ref{def1} if and only if $\varrho$ is diagonal in the orthonormal basis \eqref {6}. The latter property, in turn, is equivalent to vanishing negativity; that is,
\begin{equation}
\mathcal{N}_{\mathcal{A},\mathcal{B}}=(\Vert\varrho^{T_{\mathcal{A}}}\Vert_1-1)/2=0,
\end{equation}
where $\varrho^{T_{\mathcal{A}}}$ denotes the partial transposition with respect to mode $\mathcal{A}$
\begin{equation}
\langle k,N-k\vert\varrho^{T_{\mathcal{A}}}\vert \ell,N-\ell\rangle=\langle \ell,N-k\vert\varrho\vert k,N-\ell\rangle,
\end{equation}
with $\vert \ell,N-\ell\rangle$ being eigenvectors of $\mathrm{a}^\dag \mathrm{a}$ and $\mathrm{b}^\dag \mathrm{b}$, and $\Vert X\Vert_1=\mathrm{Tr}[\sqrt{X^{\dag}X}]$.
\end{proposition}

The main technical ingredients for the proof of is that the only $N$ boson two-mode pure states that are $(\mathcal{A},\mathcal{B})$-separable are the vector states in Eq.~\eqref{6}. Thereby, by convexity, the only $(\mathcal{A},\mathcal{B})$-separable mixed states must be their convex combinations. Furthermore,
\begin{equation}
(\varrho^{T_{\mathcal{A}}})^\dag\varrho^{T_{\mathcal{A}}}=\sum_{k,\ell=0}^N|\varrho_{k\ell}|^2\vert \ell,N-k\rangle\langle \ell,N-k\vert.
\end{equation}

Looking at the negativity of $\varrho(t)$ in Eq.~\eqref{3}, one then finds that
\begin{equation}
\mathcal{N}_{\mathcal{A},\mathcal{B}}(\varrho(t))=\frac{1}{2}\sum_{k\neq\ell}{\rm e}^{-\gamma t(k-\ell)^2/2}\,|\varrho_{k\ell}|\leqslant {\rm e}^{-\gamma t/2}\mathcal{N}_{\mathcal{A},\mathcal{B}}(\varrho)
\label{4}
\end{equation}
decreases more than exponentially; therefore, the entanglement relative to $(\mathcal{A},\mathcal{B})$ carried by $\varrho(t)$
can only be destroyed by the dissipative time evolution defined by Eq.~\eqref{1}.

As already observed, it is clear from Definition \ref{def1} that separability and entanglement are bipartition-dependent properties.
Suppose one changes the modes by a Bogolubov rotation and passes from $\mathrm{a}$ and $\mathrm{b}$ modes to the modes described by the following annihilation operators:
\begin{equation}
\mathrm{c}=(\mathrm{a}+\mathrm{b})/\sqrt{2},\quad \mathrm{d}= (\mathrm{a}-\mathrm{b})/\sqrt{2}.
\label{7}
\end{equation}
The pairs $\mathrm{c}$ and $\mathrm{d}$ generate two commuting subalgebras $\mathcal{C}$ and $\mathcal{D}$, and thus define a new local bipartition. Relative to these new modes, the $(\mathcal{A},\mathcal{B})$-separable pure states read
\begin{equation}
\vert k,N-k\rangle_{\mathcal{A},\mathcal{B}}=\frac{1}{\sqrt{2^N}}\frac{\big(\mathrm{c}^\dag- \mathrm{d}^\dag\big)^k\big(\mathrm{c}^\dag+ \mathrm{d}^\dag\big)^{N-k}}{\sqrt{k!(N-k)!}}\vert 0\rangle,
\label{8}
\end{equation}
which are linear combinations of the eigenstates
\begin{equation}
\vert k,N-k\rangle_{\mathcal{C},\mathcal{D}}=\frac{(\mathrm{c}^\dag)^{k}~(\mathrm{d}^\dag)^{(N-k)}}{\sqrt{k!(N-k)!}}\vert 0\rangle
\label{9}
\end{equation}
of $\mathrm{c}^\dag\mathrm{c}+\mathrm{d}^\dag\mathrm{d}$. Hence, the $(\mathcal{A},\mathcal{B})$-separable states are not necessarily $(\mathcal{C},\mathcal{D})$-separable. In fact, these states have nonvanishing negativity computed with respect to the orthonormal basis $\{\vert k,N-k\rangle_{\mathcal{C},\mathcal{D}}\}_{k=0}^N$ in the $N$-boson sector.

Starting with a $(\mathcal{C},\mathcal{D})$-separable state, e.g., $\varrho=|N,0\rangle_{\mathcal{C},\mathcal{D}}\langle N,0|$, for which $\mathcal{N}_{\mathcal{C},\mathcal{D}}(\varrho)=0$, one can get $(\mathcal{C},\mathcal{D})$-entanglement out of the dissipative time evolution. Here the state evolves into
\begin{equation}
\varrho(t)=\frac{1}{2\sqrt{\pi}}\int_{-\infty}^{+\infty}{\rm d}u~{\rm e}^{-u^2/4}
\vert\xi_t\rangle_{\mathcal{C},\mathcal{D}} \langle\xi_t\vert \equiv \Gamma_t[\varrho],
\label{10}
\end{equation}
where
\begin{equation}
\vert\xi_t\rangle_{\mathcal{C},\mathcal{D}} =\frac{1}{\sqrt{N!}} \mathrm{e}^{-i\sqrt{\gamma t/2}u J_z}(\mathrm{c}^\dag)^N\vert 0\rangle=
\frac{1}{\sqrt{N!}}\big(\sqrt{\xi_t}\mathrm{c}^\dag+i\sqrt{1-\xi_t}\mathrm{d}^\dag\big)^N\vert 0\rangle,
\label{11}
\end{equation}
with $\xi_t \equiv\cos^2(u\sqrt{\gamma t/2})$. Hence, $\mathcal{N}_{\mathcal{C},\mathcal{D}}(\varrho(t))>0$. The reason is that the rotations appearing in the solution are local with respect to $(\mathcal{A},\mathcal{B})$, that is, they split into product of operators, one belonging to $\mathcal{A}$, the other to $\mathcal{B}$,
\begin{equation}
\mathrm{e}^{-i\sqrt{\gamma t/2} J_z}=\mathrm{e}^{-i\sqrt{\gamma t/2}\mathrm{a}^\dag\mathrm{a}/2}\mathrm{e}^ {i\sqrt{\gamma t/2}\mathrm{b}^\dag\mathrm{b}/2},
\label{12}
\end{equation}
which of course is not possible in the new mode picture
\begin{equation}
\mathrm{e}^{-i\sqrt{\gamma t/2} J_z}=\mathrm{e}^{-i\sqrt{\gamma t/2}(\mathrm{c}^\dag\mathrm{d}+\mathrm{d}^\dag\mathrm{c})/2},
\label{13}
\end{equation}
that is, the rotation is nonlocal with respect to $(\mathcal{C},\mathcal{D})$. This implies that the noise introduces nonlocal correlation between certain modes while it destroys such correlation among other modes.

%%%%%%%%%%%%%%%%%%%%%%%%%%%%%%%%%%%%%%%%%%%%%%%%%%%%%%%%%%%%%%%
\section{$N$ two-mode boson metrology}

\subsection{Unitary estimation}

From quantum metrological arguments, one knows that certain entangled states of $N$-qubit systems can be used to achieve estimation of some parameters with errors scaling as $O(1/N)$ rather than the standard (classical) shot-noise limit $O(1/\sqrt{N})$. If the parameter to be measured is the angle $\alpha$ entering a unitary rotation of the state $\varrho$ generated by the total spin component $J_z$ (i.e., through rotating the state $\varrho$ into $\mathrm{e}^{-i\alpha J_z}\varrho \mathrm{e}^{i\alpha J_z}$), one can assess whether entanglement is metrologically useful by means of the QFI relative to $\varrho$ and $J_z$, $\mathcal{F}[\varrho,J_z]$. The QCRB here reads as
\begin{equation}
\delta \alpha\geqslant \frac{1}{\sqrt{\mathcal{F}[\varrho,J_z]}}.
\label{QCR}
\end{equation}
In order to beat the shot-noise limit, $\mathcal{F}[\varrho,J_z]$ must scale faster than $N$ with respect to the number of particles. For pure states, the QFI amounts to the mean square error relative to the generator \cite{meter}
\begin{equation}
\mathcal{F}[\vert\Psi\rangle,J_z]=4\big(\langle\Psi\vert J^2_z\vert\Psi\rangle-\langle\Psi\vert J_z\vert\Psi\rangle^2\big).
\label{QFI}
\end{equation}
While for distinguishable qubits $\mathcal{F}\big[\vert\Psi\rangle,J_z\big]\leqslant N$ unless $\vert\Psi\rangle$ is a certain entangled, for $N$ two-mode bosons $\mathcal{F}[\vert\Psi\rangle,J_z]> N$ can be achieved also by separable $\vert\Psi\rangle$. For example, consider the case of $N$ entangled two-mode boson states $\vert k,N-k\rangle_{\mathcal{C},\mathcal{D}}$ of the form \eqref{9}, which are separable with respect to the $(\mathcal{C},\mathcal{D})$ bipartition. Noting that $\displaystyle J_z=(\mathrm{c}\mathrm{d}^\dag+\mathrm{c}^\dag\mathrm{d})/2$, Eq.~(\ref{QFI}) yields
\begin{equation}
\mathcal{F}\big[\vert k,N-k\rangle_{\mathcal{C},\mathcal{D}},J_z\big]=N+2k(N-k).
\label{QFIBosons}
\end{equation}
It then follows that $\mathcal{F}[\vert k,N-k\rangle_{\mathcal{C},\mathcal{D}},J_z]>N$, unless $k=0,N$. Hence, except these two latter cases, a relative improvement over the shot-noise limit can arise, which for balanced states $k=N/2$, beats the shot-noise limit as the QFI scales as $N^2$. This improvement, despite separability of $\vert k,N-k\rangle_{\mathcal{C},\mathcal{D}}$ can be attributed to the nonlocality of the transformation $\mathrm{e}^{-i\gamma J_z}$ with respect to the bipartition $(\mathcal{C},\mathcal{D})$.

We have seen in the previous section that, under the dissipative dynamics in Eq. \eqref{10}, the $(\mathcal{C},\mathcal{D})$-separable state $\vert N,0\rangle_{\mathcal{C},\mathcal{D}}$ becomes entangled. One may thus wonder whether such dissipatively-generated entanglement may help attain a sub-shot noise error at some later time $t>0$; that is, whether $\mathcal{F}[\varrho(t),J_z]>N$ [where $\varrho(t)=\Gamma_t[\vert N,0\rangle_{\mathcal{C},\mathcal{D}}\langle N,0\vert]$] if one starts initially from $\vert N,0\rangle_{\mathcal{C},\mathcal{D}}$ [for which $\mathcal{F}[\vert N,0\rangle_{\mathcal{C},\mathcal{D}},J_z]=N$]. The answer is negative because the QFI cannot increase under trace-preserving, completely positive maps. Here $\Gamma_t$ constitute a semigroup, namely $\Gamma_{t+s}=\Gamma_t\circ\Gamma_s=\Gamma_s\circ\Gamma_t$ for any $s,t\geqslant 0$. This property implies that
\begin{equation}
\mathcal{F}[\varrho(t),J_z]=\mathcal{F}\big[\Gamma_{t-s+s}[\varrho],J_z\big]=\mathcal{F}\big[\Gamma_{t-s}[\varrho(s)],J_z\big]\leqslant \mathcal{F}[\varrho(s),J_z].
\label{14}
\end{equation}
Thus, if at time $s$, we have $\mathcal{F}[\varrho(s),J_z]\leqslant N$, $\mathcal{F}$ will remain bounded by $N$ in the course of time despite the entanglement generated by the time evolution.

%%%%%%%%%%%%%%%%%%%%%%%%%%%%%%%%%%%%%%%%%%%%%%%%%%%%%%%%%%%%%%%
\subsection{Dissipative estimation}

We now change the perspective and consider the approach wherein the parameter to be estimated is in the very dissipative quantum time evolution itself. Namely, instead of unitarily rotating with $\mathrm{e}^{-i\alpha J}$ a time-evolving $\varrho(t)$ in order to measure the parameter $\alpha$, we consider the parameter $\gamma$ to be estimated as inscribed in the time evolving state $\varrho(t)$ itself. In the case of a unitary time evolution with the Hamiltonian generator $H_{\gamma}=\gamma H$,
\begin{equation}
\varrho\mapsto \mathrm{e}^{-it\gamma H}\varrho \mathrm{e}^{it\gamma H},
\end{equation}
we could try to estimate the time-scale parameter $\gamma$. Since pure states are mapped into pure states, starting off with $\varrho=\vert\Psi\rangle\langle\Psi\vert$ one can use Eq.~(\ref{QFI}) with $J_z$ replaced by $t H$.

Notice that, if $H=J_z=(\mathrm{c}^\dag\mathrm{d}+\mathrm{c}\mathrm{d}^{\dag})/2$, the ensuing unitary dynamics would be generically entangling a $(\mathcal{C},\mathcal{D})$-separable state as $\vert N,0\rangle_{\mathcal{C},\mathcal{D}}$; however, for the QFI, Eq.~(\ref{QFI}) would give the linear scaling $\displaystyle \mathcal{F}[\vert N,0\rangle_{\mathcal{C},\mathcal{D}}, t J_z]=t^2\,N$. Suppose, instead, the time evolution is described by a semigroup of CPT maps generated by a master equation of the form $\partial_t\varrho(t)=L_{\gamma}[\varrho(t)]$ with the Lindblad-type generator $L_{\gamma}$, where again the parameter $\gamma$ to be estimated rescales time $L_{\gamma}=\gamma L$.

Consider, for example, the master equation \eqref{ME}. This evolution is the result of the coupling of the system of interest to an external environment which is then traced away according to a procedure called weak-coupling limit. The parameter $\gamma$ thus is the strength of the decoherence effects due to the coupling to the environment.

In order to deal with such cases, we follow the framework outlined in Ref.~\cite{dissipative-Cramer-Rao} for dissipative metrology and QCRB---for other relevant studies, see, e.g., Refs.~\cite{Guta:NatureC}. One can recast the density matrix formalism into a vector-like formalism, as follows. Let $\varrho(t)$ be an $n\times n$ density matrix and $\{\vert k\rangle\}_{k=1}^n$ be an orthonormal basis in $\mathbb{C}^n$, then one associates to $\varrho(t)$ the $n^2$-dimensional vector $\Ket{\tilde{\varrho}(t)}$ whose components are the density matrix entries $\varrho_{k\ell}(t)=\langle k\vert\varrho(t)\vert \ell\rangle$ with respect to the chosen orthonormal basis. Actually, this can be done for all matrices $X\in M_n(\mathbb{C})$ so that the matrix algebra is mapped into an $n^2$-dimensional Hilbert space with the scalar product defined by
\begin{equation}
\BraKet{X}{Y}=\mathrm{Tr}[X^{\dag}Y].
\label{scalprod}
\end{equation}
Furthermore, all linear operators on the algebra are represented as $n^2\times n^2$ matrices on such a Hilbert space; denoting by $\mathcal{L}$ the matrix corresponding to the generator $L$, the dissipative time evolution on density matrices can be recast in the following vectorial form:
\begin{equation}
\varrho(t)=\mathrm{e}^{tL_{\gamma}}[\varrho]\longmapsto \Ket{\tilde{\varrho}(t)}={\rm e}^{t\gamma\mathcal{L}}\Ket{\tilde{\varrho}}.
\label{vectrep}
\end{equation}
Notice that the vector state $\Ket{\tilde{\varrho}}$ is not normalized and that the matrix $\mathcal{L}$ is in general non-Hermitian. However, the normalization $\Ket{\varrho}=\Ket{\tilde{\varrho}}/\sqrt{{\rm Tr}[\varrho^2]}$ is preserved by trace-preserving time evolution so that Eq.~\eqref{vectrep} extends to normalized vectors.
In this extended description, the QFI associated to the vector $\Ket{\varrho(t)}$ and the operator $t\mathcal{L}$ is given by
\begin{eqnarray}
\mathcal{F}\big[\Ket{\varrho(t)},t\mathcal{L}\big]=4t^2\big(\Bra{\varrho(t)}\mathcal{L}^\dag\mathcal{L}\Ket{\varrho(t)}
\,-\,\left|\Bra{\varrho(t)}\mathcal{L}\Ket{\varrho(t)}\right|^2\big).
\label{QFIext}
\end{eqnarray}
In addition, the dissipative QFI $\mathcal{F}\big[\Ket{\varrho(t)},t\mathcal{L}\big]$ can be used to bound the standard QFI $\mathcal{F}\big[\varrho(t),tL\big]$ as follows:
\begin{equation}
\mathcal{F}\big[\Ket{\varrho(t)},t\mathcal{L}\big]\frac{\mathrm{Tr}[\varrho^2(t)]}{4\lambda_{\max}\big(\varrho(t)\big)}< \mathcal{F}\big[\varrho(t),tL\big]<
\big(\mathcal{F}\big[\Ket{\varrho(t)},t\mathcal{L}\big]+g(t)\big)\frac{\mathrm{Tr}[\varrho^2(t)]}{4\lambda_{\min}\big(\varrho(t)\big)},
\label{fish-ineq}
\end{equation}
in which $g=4\big(\mathrm{Tr}[\varrho^2 G]/\mathrm{Tr}[\varrho^2]\big)^2$, with $G$ being the symmetric logarithmic derivative \cite{dissipative-Cramer-Rao}.

\begin{remark}
\label{rem}
Note that the inverse of the QFI $\mathcal{F}[\varrho(t),tL]$ itself provides a lower bound for the estimation error $(\delta \gamma)^2$. Thus finding a lower bound for $\mathcal{F}[\varrho(t),tL]$---as in Eq. (\ref{fish-ineq})---would in turn provide an upper bound for the lower bound of $(\delta \gamma)^2$. This upper bound hence can only be a more conservative estimate for the error $(\delta \gamma)^2$; that is, the true error might be even less than this crude estimate. In particular, if one finds an $O(1/N^{\beta})$ scaling for $\delta \gamma$ from this analysis, it is evident that the scaling given from the analysis of $\mathcal{F}[\varrho(t),tL]$ cannot be of $O(1/N^{\beta-\eta})$, for some $\eta>0$. This argument justifies that our error scaling analysis based on the left hand side of Eq.~(\ref{fish-ineq}) is safe and rigorous but perhaps (too) conservative.
\end{remark}

We apply the dissipative approach first to the case where the time evolution is unitary and generated by the Hamiltonian $H=\gamma J_z$, to compare it with the standard approach within which we have seen that no improvement on the shot-noise limit is possible starting from the $(\mathcal{C},\mathcal{D})$-separable state $\vert N,0\rangle_{\mathcal{C},\mathcal{D}}$. Next, we will consider the case of the dissipative time evolution Eq.~\eqref{3}, and study whether the entanglement generated acting on $\vert N,0\rangle_{\mathcal{C},\mathcal{D}}$ is useful to go beyond the shot-noise limit in the estimation of the parameter $\gamma$. Given the master equation
\begin{equation}
\partial_t\varrho(t)=-i[\gamma J_z,\varrho(t)],
\label{ham}
\end{equation}
and using the representation given by the basis vectors in Eq.~\eqref{6}, one obtains
\begin{equation}
\partial_t\varrho_{k\ell}(t)=-i\gamma (k-\ell)\,\varrho_{k\ell}(t)\Longrightarrow \varrho_{k\ell}(t)=\mathrm{e}^{-it\gamma(k-\ell)}\varrho_{k\ell}\ .
\label{ham1}
\end{equation}
The time evolving $N\times N$ density matrix $\varrho(t)$ corresponds to the $(N+1)^2$-dimensional normalized vector $\Ket{\varrho(t)}$, with the components $\varrho_{k\ell}(t)/\sqrt{\mathrm{Tr}[\varrho^2(t)]}$.
Moreover, from Eq.~\eqref{ham} one finds that the generator $L$ amounts to a diagonal matrix
\begin{equation}
\mathcal{L}_{k\ell,pq}=-i(k-\ell)\delta_{kp}\delta_{\ell q}.
\label{MatGenHam}
\end{equation}
Thus, Eq.~\eqref{QFIext} reads
\begin{equation}
\mathcal{F}\big[\Ket{\varrho(t)},t\mathcal{L}\big]=\frac{t^2}{\mathrm{Tr}[\varrho^2(t)]}\left(\sum_{k,\ell=0}^N \left|\varrho_{k\ell}(t)\right|^2(k-\ell)^2-\Big(
\sum_{k,\ell=0}^N \left|\varrho_{k\ell}(t)\right|^2(k-\ell)\Big)^2\right)=\frac{t^2}{\mathrm{Tr}[\varrho^2]}\,\sum_{k,\ell=0}^N \left|\varrho_{k\ell}\right|^2(k-\ell)^2.
\label{QFIext2Ham}
\end{equation}
Taking $(\mathcal{C},\mathcal{D})$-separable state as the initial condition
\begin{equation}
\vert N,0\rangle_{\mathcal{C},\mathcal{D}}=\frac{1}{\sqrt{N!}}\left(\frac{\mathrm{a}^\dag+\mathrm{b}^\dag}{\sqrt{2}}\right)^N\vert 0\rangle=
\frac{1}{\sqrt{2^N}}\sum_{k=0}^N\sqrt{N\choose k}\vert k,N-k\rangle_{\mathcal{A},\mathcal{B}}
\label{incond}
\end{equation}
yields $\mathrm{Tr}[\varrho^2]=1$, whence
\begin{equation}
\mathcal{F}\big[\Ket{\varrho(t)},t\mathcal{L}\big]=\frac{t^2}{2^{2N}}\,\sum_{k,\ell=0}^N (k-\ell)^2\, {N\choose k}\, {N\choose \ell}=\frac{t^2}{2}\,N .
\label{QFIext2Ham2}
\end{equation}
This is exactly as in the standard formalism, despite the $(\mathcal{C},\mathcal{D})$-nonlocality of the time evolution generated by $J_z$, no improvement on the linear scaling of QFI with $N$ can be achieved starting with the $(\mathcal{C},\mathcal{D})$-separable state $\vert N,0\rangle_{\mathcal{C},\mathcal{D}}$.

We now take the same initial state $\vert N,0\rangle_{\mathcal{C},\mathcal{D}}$ but this time let it evolve according to the dissipative time evolution generated by the Lindblad equation \eqref{ME}. The $(N+1)^2$-dimensional vector $\Ket{\varrho(t)}$ has components $\varrho_{k\ell}(t)/\sqrt{\mathrm{Tr}[\varrho^2(t)]}$, where
\begin{equation}
\varrho_{k\ell}(t)=\frac{1}{2^N}\sqrt{{N\choose k} {N\choose \ell}}\mathrm{e}^{-\gamma t(k-\ell)^2/2}\ .
\label{rho-components}
\end{equation}
Moreover, the generator
$L$ amounts to a diagonal matrix
\begin{equation}
\mathcal{L}_{k\ell,pq}=-\frac{1}{2}(k-\ell)^2\delta_{kp}\delta_{\ell q}.
\label{MatGen}
\end{equation}
Thus, Eq.~\eqref{QFIext} reads
\begin{eqnarray}
\mathcal{F}\big[\Ket{\varrho(t)},t\mathcal{L}\big]&=&t^2\left(\sum_{k,\ell=0}^N \left|\varrho_{k\ell}(t)\right|^2(k-\ell)^4-\Big(
\sum_{k,\ell=0}^N \left|\varrho_{k\ell}(t)\right|^2(k-\ell)^2\Big)^2\right)
\label{QFIext2}
\\
\mathcal{F}\big[\Ket{\varrho(t)},t\mathcal{L}\big]&=&\frac{t^2}{2^{2N}}\left(\frac{1}{\mathrm{Tr}[\varrho^2(t)]}\sum_{k,\ell=0}^N {N\choose k} {N\choose \ell}(k-\ell)^4\mathrm{e}^{-\gamma t(k-\ell)^2}\right. \nonumber\\
&&-\frac{1}{2^{2N}}\frac{1}{(\rm{Tr}[\varrho^2(t)])^2}
\left.\left(\sum_{k,\ell=0}^N {N\choose k}{N\choose \ell}(k-\ell)^2\mathrm{e}^{-\gamma t(k-\ell)^2}\right)^2\right).
\label{QFIext3}
\end{eqnarray}

In order to extract the scaling behavior vs. $N$ from the above expression at a generic time $t\geqslant 0$, one has to appeal to numerical simulations. However, at least for small times one can proceed analytically and obtain the first relevant term of the order $t^2$. Consider first $\mathrm{Tr}[\varrho^2(t)]$; by using Eq.\eqref{expon} it can be recast as
\begin{equation}
\mathrm{Tr}[\varrho^2(t)]=2^{-2N}\sum_{k,\ell=0}^N{N\choose k} {N\choose \ell}\mathrm{e}^{-\gamma t(k-\ell)^2}=\int_{-\infty}^{+\infty}\mathrm{d}u\,\frac{\mathrm{e}^{-u^2}}{\sqrt{\pi}}\, \cos^{2N}(u\sqrt{\gamma t})\ .
\label{expon1}
\end{equation}
By setting $\mu=\gamma t$ and $I(\mu)=\mathrm{Tr}[\varrho^2(t)]$, the sums contributing to Eq. \eqref{QFIext3} can be evaluated as
\begin{eqnarray}
\label{expon2a}
I_1(\mu)&=&\sum_{k,\ell=0}^N {N\choose k}{N\choose \ell}(k-\ell)^2\mathrm{e}^{-\mu(k-\ell)^2}=-\frac{\mathrm{d}I(\mu)}{\mathrm{d}\mu}=\frac{N2^{2N}}{\sqrt{\mu}}\int_{-\infty}^{+\infty}\mathrm{d}u\,\frac{\mathrm{e}^{-u^2}}{\sqrt{\pi}}
u\cos^{2N-1}(u\sqrt{\mu})\,\sin(u\sqrt{\mu})\\
\\
\nonumber
I_2(\mu)&=&\sum_{k,\ell=0}^N {N\choose k}{N\choose \ell}(k-\ell)^4\mathrm{e}^{-\mu(k-\ell)^2}=\frac{\mathrm{d}^2I(\mu)}{\mathrm{d}\mu^2}=\frac{N2^{2N}}{2\mu\sqrt{\mu}}\int_{-\infty}^{+\infty}\mathrm{d}u\,\frac{\mathrm{e}^{-u^2}}{\sqrt{\pi}}
u\cos^{2N-1}(u\sqrt{\mu})\,\sin(u\sqrt{\mu})\\
\nonumber
&&+
\frac{N(2N-1)2^{2N}}{2\mu}\int_{-\infty}^{+\infty}\mathrm{d}u\,\frac{\mathrm{e}^{-u^2}}{\sqrt{\pi}}
u^2\cos^{2N-2}(u\sqrt{\mu})\,\sin^2(u\sqrt{\mu})
\\
&&-
\frac{N2^{2N}}{2\mu}\int_{-\infty}^{+\infty}\mathrm{d}u\,\frac{\mathrm{e}^{-u^2}}{\sqrt{\pi}}
u^2\cos^{2N}(u\sqrt{\mu}).
\label{expon2b}
\end{eqnarray}
The limits of the above two expressions for $\mu\to0$ are
\begin{equation}
I_1(0)=\frac{N\,2^{2N}}{2}\ ,\qquad I_2(0)=\frac{2^{2N}3N(2N-1)}{8},
\end{equation}
whereby to the leading order in $t$,
\begin{equation}
\mathcal{F}\big[\Ket{\varrho(t)},t\mathcal{L}\big]=\frac{t^2N^2}{2}\left(1-\frac{3}{8N}\right)+o(t^2),
\label{expon3}
\end{equation}
thus exhibiting a scaling behavior that beats the shot-noise limit.

As explained in Remark \ref{rem}, it is important to study the scaling behavior of the lower bound in Eq. \eqref{fish-ineq}. In order to do that numerically, one argues as follows. From Eq.~(\ref{rho-components}), it is seen that $\varrho(t)$ is an almost diagonal matrix, since its off-diagonal elements exponentially decay with the distance from the diagonal. We employ this observation and approximate the largest eigenvalue of the matrix with its largest diagonal element, i.e.,
\begin{equation}
\lambda_{\max}\big(\varrho(t)\big)\approx \frac{1}{2^N} {N\choose N/2}.
\label{lambda_max}
\end{equation}
Since this approximation works best only for large $t$, to alleviate this problem, we use the fact that $\lambda_{\max}\leqslant 1$, and rewrite the left hand side of Eq.~(\ref{fish-ineq}) as follows:
\begin{equation}
\mathcal{F}\big[\varrho(t),tL] >\mathcal{F}\big[\Ket{\varrho(t)},t\mathcal{L}\big]\mathrm{Tr}[\varrho^2(t)]/4.
\label{fish-ineq-2}
\end{equation}

\begin{figure}[h]
\includegraphics[scale=.7]{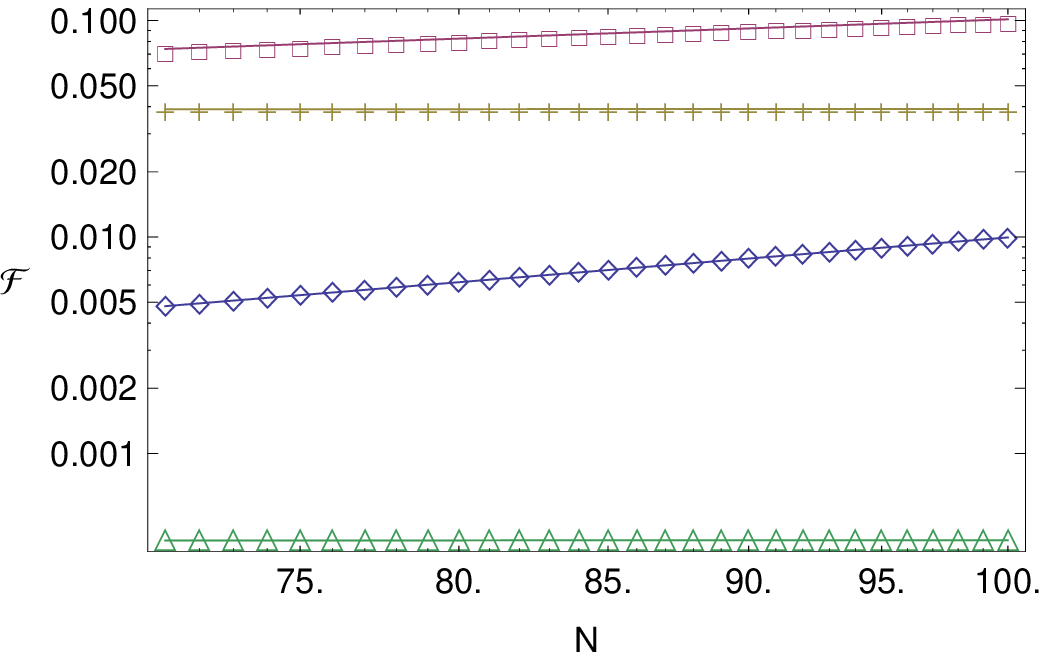} \hskip13mm \includegraphics[scale=.7]{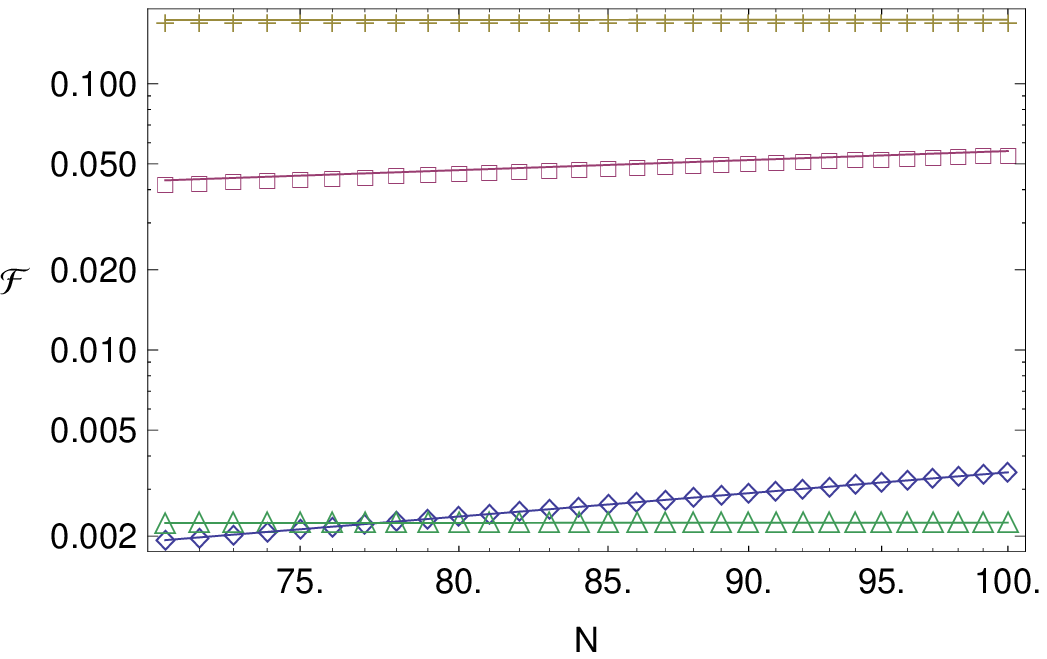}
\caption{(Color online). Scaling of the dissipative QFI vs. $N$ at $t=0.001$ ($\Diamond$), $0.01$ ($\square$), $1$ ($+$), and $5$ ($\triangle$) in the log-log plot. We have fitted a function of the form $\alpha N^{\beta}$. Left: the lower bound of Eq.~(\ref{fish-ineq}). The scaling parameters $(\alpha,\beta)$ are, respectively, $(5.36239\times10^{-7}, 2.13436)$, $(0.0014597, 0.920703)$, $(0.0374129,  0.00929577)$, and $(0.000379478,0.010242)$. Right: the QFI as presented in Eq.~(\ref{QFIext3}). $(1.30898\times10^{-6},1.71177)$, $(0.00185645,0.739184)$, $(0.167394,0.00828191)$, and $(0.00212922,0.0117239)$.}
\label{scaling}
\end{figure}

Figure~\ref{scaling} shows the scaling behavior of the dissipative QFI vs. $N$ for different values of $t$. It is seen that for short times (such as $t=0.001$) the dissipative QFI can show a scaling of the form $O(N^{\beta})$ with $\beta>1$ for the lower bound of the QFI calculated from Eq.~(\ref{fish-ineq}) [indeed for $t=0.001$ our numerics indicates that $\beta\approx2$, in agreement with the analytical result in Eq.~\eqref{expon3}], whereas for larger times as expected for Markovian dynamics the QFI vanishes exponentially.

Therefore, in a scenario where we aim to measure the typical life-time of a dissipative dynamics dynamics, our result indicates that the dissipatively generated entanglement may improve metrology at small enough evolution times.

%%%%%%%%%%%%%%%%%%%%%%%%%%%%%%%%%%%%%%%%%%%%%%%%%%%%%%%%%%%%%%%
\section{Summary and Conclusions}

How estimation error scales with the number of probes in a metrological task is of paramount importance in quantum estimation. Exploiting various resources such as entanglement among probes, using scenarios with manybody interaction, or nonclassicality have all proven useful under some conditions for quantum estimation. It is thus a natural question whether interaction with an environment may also induce resources useful for metrology. At first glance, it may be argued that interactions with ambient environment is typically dissipative and results in decoherence, whence such interactions are mostly destructive for estimation. This, however, needs a more careful analysis, since it is now believed that open quantum dynamics, under certain circumstances, may also offer advantages for various quantum tasks.

Bearing this in mind, we aimed to analyze whether entanglement generated through a dissipative dynamics may be exploited to provide a partially better error scaling with probe system size in the estimation of a parameter of the very dissipative dynamics. We first considered an open quantum system described by the Markovian dynamics of (collective) dephasing acting on an $N$ particle two-mode bosonic system, and analytically found the exact form of the evolution of the states. Next, we reviewed the concept of entanglement in quantum systems of identical particles. We showed that this dynamics generated entanglement between modes of our system. Using an open system framework for (dissipative) quantum Fisher information and the quantum Cram\'er-Rao bound, we analyzed a scenario in which the typical life-time of the dissipative dynamics was to be estimated. Our analysis indicated that, under some conditions and for small enough evolution times, a better scaling of $O(N^{\beta})$, with $\beta>1$, shows up in estimates of the quantum Fisher information.

%%%%%%%%%%%%%%%%%%%%%%%%%%%%%%%%%%%%%%%%%%%%%%%%%%%%%%%%%%%%%%%
\textit{Acknowledgments.}---ATR acknowledges partial financial support from Sharif University of Technology's Office of Vice-President for Research.

%%%%%%%%%%%%%%%%%%%%%%%%%%%%%%%%%%%%%%%%%%%%%%%%%%%%%%%%%%%%%%%
%%%%%%%%%%%%%%%%%%%%%%%%%%%%%%%%%%%%%%%%%%%%%%%%%%%%%%%%%%%%%%%

%%%%%%%%%%%%%%%%%%%%%%%%%%%%%%%%%%%%%%%%%%%%%%%%%%%%%%%%%%%%%%%

\end{document}